\documentclass{article}
\usepackage[ansinew]{inputenc}

\usepackage[dvips]{graphicx}

\usepackage{url}
\usepackage{amsmath}

\usepackage{color}
 \usepackage{setspace}

 \usepackage[square,numbers, super, sort]{natbib}

\title{Bayesian joint models for longitudinal and survival data}
 \author{%
  \large Carmen Armero\ \\[1ex] 
\large   Department of Statistics and Operations Research, Universitat de Val\`encia \\
C/ Doctor Moliner 50, Burjassot. 46100 Val\`encia, Spain\\
}
\date{ }

\begin{document}

\maketitle

\begin{abstract}
   This paper takes a quick look at Bayesian joint models (BJM) for longitudinal and survival data. A general formulation for BJM is examined in terms of the sampling distribution of the longitudinal and survival processes, the conditional distribution of the random effects and the prior distribution.  Next a basic BJM defined in terms of a mixed linear model and a Cox survival regression models is discussed and some extensions   and other Bayesian  topics are briefly outlined.\vspace*{-0.3cm}\\

\end{abstract}

\section{Introduction}

Longitudinal data are observations of one or more variables measured over time
of each of the individuals in the study.
They include observations between and within individuals that allow  the assessment of general patterns of the target population   as well as specific individual characteristics.
These data are multivariate, clustered, and repeated measures. Longitudinal data   of each individual could be understood as a time series. Panel data are longitudinal data in economic scenarios.

Survival times   measure follow-up times from a defined starting point to the occurrence of a given event or endpoint of interest. Standard statistical techniques cannot  be applied to survival data because
they are subject to censoring and/or truncation schemes that usually do not provide complete experimental information.

Joint modeling of longitudinal and survival data is an increasing and productive area of statistical research that examines the association between longitudinal and survival processes. It enhances survival interests with the inclusion of internal time-dependent covariates  assessed through longitudinal models as well as longitudinal objectives by allowing for the inclusion of non-ignorable dropout mechanisms through survival tools (Verbeke and Davidian, 2009; Ye and Yu, 2014). Joint models were introduced during the 1990s as a consequence of the research about the human immunodeficiency virus infection and acquired immune deficiency syndrome (HIV/AIDS), and cancer studies. Since then, they have been applied to a great variety of studies, mostly in  epidemiological and biomedical areas.

 The two key elements of the Bayesian reasoning   (with regard to the frequentist one)  are the conception of probability, that allows to measure uncertainty associated to parameters, models, hypotheses, missing data, etc. in probabilistic terms, and the use of Bayes' theorem to sequentially update probabilities as more relevant information is obtained. As a result, Bayes inference offers a wide and attractive framework to joint models of longitudinal and survival analysis: posterior inferences for any outcome of interest depending on the parameters  that makes unnecessary asymptotic approximations, a simple framework to easily incorporate historical data into the inferential process,   or prediction of observable quantities   directly assessed in probabilistic terms among others (Ibrahim et al, 2001).

 The literature on joint models  of longitudinal and survival data is enormous (Rizopoulos, 2012) is an excellent text within the frequentist methodology). Bayesian   literature is not as extensive  but too numerous to be cited here. References in this paper are only a very small part of all those that should appear  in any more extensive work on  the area. In any  case, a reminder of  the early works on the topic  (Faucett and Thomas, 1996; Ibrahim et al.,  2001;  Wang and Taylor, 2001) is always more than convenient  to have a historical perspective of the beginning research in the subject.

 \section{Bayesian joint models}

\subsection{General formulation}

A BJM for longitudinal and survival data   is a  joint probability distribution for the observable longitudinal ($y$) and  survival variables ($s$) as well as for the corresponding subject-specific random effects vector ($\boldsymbol b$) and the parameters and hyperparameters ($\boldsymbol \theta$) of the model.
This density could be generally expressed as

\begin{equation}
p(y, s,  \boldsymbol b, \boldsymbol  \theta) = p(y, s \mid  \boldsymbol b,  \boldsymbol \theta ) \, p( \boldsymbol b  \mid  \boldsymbol \theta ) \, p( \boldsymbol \theta ),
\label{eqn:jointmodel}
\end{equation}

\noindent where  $p(y, s \mid \boldsymbol b, \boldsymbol \theta)$ is the sampling distribution   of $(y,s)$ given $(\boldsymbol b, \boldsymbol \theta)$,   $p(\boldsymbol b \mid \boldsymbol \theta)$  the conditional distribution of the random effects given $\boldsymbol \theta$, and $p(\boldsymbol \theta)$ the prior distribution for $\boldsymbol \theta$. Note that the subsequent formulation in the frequentist framework would not include the vector $\boldsymbol \theta$  and therefore its definition  would be only written in terms  of   $p(y, s, \boldsymbol b)$, $p(y, s \mid \boldsymbol b)$ and $p(\boldsymbol b)$.

There are different proposals for the specification of $p(y, s \mid  \boldsymbol b, \boldsymbol \theta)$. All them    provide a  wide framework of the relationship  between the longitudinal and the survival processes which facilitates the modeling into longitudinal and survival submodels with various types of connectors between them (Verbeke and Davidian 2009). The most popular structures for that relation are the so-called  conditional models,  share-parameter models, random-effects models, and the joint latent class models. Copula based models   have also addressed that issue recently (Li et al., 2019).

Conditional models were the first to be used in the initial studies on the subject.  They express the joint sampling distribution of the longitudinal and survival process as follows:
\begin{align*}
p(y, s \mid  \boldsymbol b, \boldsymbol  \theta) = \, & p(y \mid s, \boldsymbol b, \boldsymbol \theta) \, p(s \mid  \boldsymbol b, \boldsymbol \theta)\,\,\,\mbox{(selection models)}\\
                                                 = \, & p(s \mid y, \boldsymbol b, \boldsymbol \theta) \, p(y \mid  \boldsymbol b, \boldsymbol \theta)\,\,\, \mbox{(pattern-mixture models).}
\end{align*}

Selection models are used when the interest of the study lies in  the survival process   while pattern-mixture models  are appropriate for longitudinal purposes.

Shared-parameter models connect the longitudinal and the time-to-event processes  by means of a common set of subject-specific random effects. This approach postulates conditional independence between the longitudinal and survival processes given the random effects and the parameters:

$$p(y, s \mid \boldsymbol b, \boldsymbol \theta ) = p(y \mid \boldsymbol b, \boldsymbol \theta) \, p(s \mid  \boldsymbol b, \boldsymbol \theta).$$

A clear disadvantage of these models is the stiffness of the correlation structure between the longitudinal and the survival processes. The random-effects approach (Henderson \textit{et al.}, 2000) has the same design that the shared-parameter models but allows for more flexibility (and thus more complexity) for the connection between the survival and the longitudinal processes, enabling that a part of the random effects associated to both processes are not common.

The joint latent class model (Proust-Lima et al., 2014) is based on finite mixtures:   heterogeneity among the individuals is classified into a finite number of homogeneous latent clusters, which share the same longitudinal trajectory and   risk function. Both processes are    conditionally independent within the subsequent latent group as follows:

$$p(y, s \mid G=g, \boldsymbol  b, \boldsymbol \theta) = p(y \mid G=g, \boldsymbol b, \boldsymbol \theta)\, p(s \mid G=g,\boldsymbol \theta),$$

\noindent where $G$ is the random variable that measures the uncertainty on the membership of each individual to each group, usually modelled by means of a multinomial logistic model.

The complete specification of a BJM includes the conditional distribution of the random effects, which can be both time-dependent and time-independent, and the choice of an a prior  distribution for $\boldsymbol \theta$. After their specification   and the obtention of data, Bayes' theorem updates the current information in terms of the posterior distribution $p(\boldsymbol b, \boldsymbol \theta \mid data)$ which allows posterior inferences for any outcome of interest   depending on $\boldsymbol b$ and/or $\boldsymbol \theta$ and it is the main element for assessing   posterior predictive distributions for new observable longitudinal and survival elements from

\begin{equation}
p(y, s \mid data)= \int\, p(y,s \mid \boldsymbol b, \boldsymbol \theta)\, p( \boldsymbol b, \boldsymbol \theta \mid data)\, \mbox{d}(\boldsymbol b, \boldsymbol \theta),
\end{equation}
or separately via the marginal posterior predictive distributions $p(y \mid data)$ and $p(s \mid data)$.

\subsection{A basic Bayesian joint model}

The basic BJM is composed of the two most paradigmatic longitudinal and survival models such as the linear mixed  model and the Cox survival regression model.

The natural territory of the linear mixed effects model is the normal distribution, whose conditional mean accounts for common population terms and individual-specific elements, the random effects, which are unique to each individual separately.  Let $y_{ij}$ denote the longitudinal  response variable  for the $i$th, $i=1, \ldots, N$ individual registered at the time  $t_{ij}$, $j=1, \ldots, n_i$. We assume the following linear mixed  model for    $\boldsymbol y_i=(y_{i1}, \ldots, y_{in_i})^{\prime}$:

\begin{equation}
(\boldsymbol y_i \mid \boldsymbol \beta, \boldsymbol b_i, \sigma^2)   \sim \mathcal{N}(\boldsymbol m_{i}=\boldsymbol X_i \boldsymbol \beta + \boldsymbol Z_i \boldsymbol b_i, \sigma^2\,\boldsymbol I),
\end{equation}

\noindent where $\boldsymbol X_i$ is a  $n_i \times q$   matrix of covariates associated to the parametric vector  $\boldsymbol \beta$, $\boldsymbol Z_i$ is a matrix of covariates associated to   random effects
$\boldsymbol b_i$, and $\boldsymbol I$ is the $n_i \times n_i$ identity matrix. Random effects $\boldsymbol b=(b_1, \ldots, b_N)^{\prime}$ are usually conditionally i.i.d. as
$(\boldsymbol b_i \mid \Sigma_b ) \sim \mathcal{N}(\boldsymbol 0,  \Sigma_b)$ where $\Sigma_b$  is the   variance-covariance matrix.

The Cox proportional hazards model   expresses
the hazard function $h_i(t)$  of the survival time $T_i$ of   individual $i$   as the product of a common
baseline hazard function, $h_0(t)$,    which determines the shape of $h_i(t)$, and an exponential term with the relevant
covariates  as follows:

\begin{equation}
h_i(t \mid \mathcal{M}_{i}(t), \boldsymbol \gamma, \alpha, h_0(t) )= h_0(t)\,\mbox{exp}\{\boldsymbol d_i^{\prime} \boldsymbol \gamma  + \alpha \, m_{i}(t)\},
\label{eqn:survival}
\end{equation}
where  $\mathcal{M}_i(t)=\{m_{i}(l), 0 \leq l \leq t\}$  represents the true longitudinal trajectory of individual $i$ up to time $t$, $\boldsymbol d_i$  is     a vector of baseline covariates associated to coefficients
$\boldsymbol \gamma$,   and $\alpha$ is the coefficient of association between the longitudinal and the survival process. Obviously, if $\alpha$   were zero the survival and the longitudinal process would be independent.

  The last element of the BJM is the prior distribution for $\boldsymbol \theta$ which includes all the parameters and hyperparameters of the longitudinal model, $\boldsymbol \beta$, $\Sigma_{b}$ and $\sigma^2$, as well as the subsequent of the survival process, $\boldsymbol \gamma$, $\alpha$ and the ones in $h_0(t)$. Prior independence is the most simple assumption for the joint prior distribution for  $\boldsymbol \theta$. Priors for $\boldsymbol \beta$, $\gamma$ and $\alpha$ are commonly elicited as   normal distributed centered at zero with   wide variances, inverse Wishart for $\Sigma_b$ and inverse gamma for $\sigma^2$  (Guo and Carlin, 2004), although options such as uniform or half-Cauchy distributions for variances seem more appropriate  (Gelman, 2006).

  Simple extensions of the basic BJM introduce complexity into the longitudinal and the survival modeling such as the distribution of the random effects,  the baseline hazard function, and the connectors between the longitudinal and the survival processes. More flexible longitudinal trajectories in terms of splines are in  Tang and Tang (2015). Kh\"{o}ler \textit{et al.} (2016) also extend this flexibility to the survival part of the BJM. Additional terms in the conditional mean $m_{i}(t)$ that account for serial correlation not explained by the random effects $\boldsymbol b$ are considered  via   Ornstein-Uhlenbeck stochastic processes  and   Brownian motions in
   Wang and Taylor (2001)  and   Armero \textit{et al.}  (2018), respectively. Multivariate $t$ or Laplace distributions have also been considered as appropriate   models for the random effects (Tian \textit{et al.}, 2016).

  The basic model for $h_0(t)$ is defined in terms of the Weibull distribution. Semi-parametric choices (Ibrahim et al., 2014) result in more flexible baseline shapes  that could be subject to regularization through prior distributions in case of   overfitting and instability. The shared-parameter model (SPM) in (\ref{eqn:survival})  is known as the trajectory model (Zhang \textit{et al.}, 2017) because the true longitudinal mean operates as a temporal covariate in the survival model. Other types of SPM connectors  include directly the random effects as covariates in the survival model, different classes of links with regard to individual subgroups, or even time-dependent slope components (Gould et al., 2015).

\section{Some extensions  and other topics}
Bayesian inference offers a  natural environment  to address complex models with sophisticated longitudinal and/or survival structures. In the case of the survival submodels, administrative
 right censoring is the most usual pattern  but also interval-censoring and left truncation  (Armero \textit{et al.}, 2018) can be considered. More complex structures include cure rate
  models (Yu \textit{et al.), 2004} or even spatial terms in the hazard function (Martins et al., 2016).  Survival submodels with more that one event of interest include, among others, recurrent
   and competing  risks models (Hu et al., 2009). BJM's with non-normal longitudinal response focused on ordinal  longitudinal processes   defined in terms of
 proportional-odds cumulative logit models and  longitudinal zero-inflated counts are in   Armero \textit{et al.} (2016) and Zhu \textit{et al.} (2018), respectively.  Luo (2014) accounts for BJM with multivariate longitudinal
binary, ordinal cumulative probabilities, and   continuous
outcomes.

 Joint models are complex models. Their practical implementation is  challenging  and consequently, an important issue in Bayesian computation. In addition to the general software for  Bayesian models, we would like to mention  the \texttt{R} packages \texttt{JMbayes} (Rizopoulos, 2016) and \texttt{bamlss} (Umlauf \textit{et al.}, 2018; Umlauf \textit{et al.}, 2019).   Furgal \textit{et al.} (2019)   is a recent review on the subject which    can currently be completed with  Niekerk \textit{et al.}  (2019),
   which explores BJM via  \texttt{R-INLA}   based on the Integrated Nested Laplace Approximations (INLA) methodology.

   Model diagnosis and model selection in BJM are very relevant issues that have been given little attention. Some papers to be quoted are   Zhu \textit{et al.} (2012), which accounts for   influence measures for carrying out  sensitivity
analysis to BJM, and  Zhang \textit{et al.} (2014)  that derives a novel decomposition of the AIC and BIC criteria into
additive components that   allow  to assess the goodness of fit for each component of
the joint model.   Huang and Dagne (2011) consider   BJM with
skew-normal distribution and measurement errors in covariates,  Armero \textit{et al.} (2016) deals with  dynamic estimation and prediction, and finally  Alvares \textit{et al.} (2020) introduces sequential Monte Carlo methods to update the posterior distribution as more information becomes available. \vspace*{0.2cm}\\

\noindent \textbf{\large References}

\begin{enumerate}
\item  D. Alvares, C. Armero, A. Forte, and N. Chopin. Sequential Monte Carlo methods in Bayesian joint models
for longitudinal and time-to-event data. \textit{Statistical Modelling},  doi.org/10.1177/1471082X20916088, 2020.
\item  C. Armero, C. Forn\'e, M. Ru\'e, A. Forte, H. Perpin\'an, G. G\'omez, and M. Bar\'e. Bayesian joint ordinal and
survival modeling for breast cancer risk assessment. \textit{Statistics in Medicine}, 35:5267-5282, 2016.
\item  C. Armero, A. Forte, H. Perpin\'an, M. J. Sanahuja, and S. Agust\'i. Bayesian joint modeling for assessing the
progression of Chronic Kidney Disease in children. \textit{Statistical Methods in Medical Research}, 27(1):298-311,
2018.
\item  C. L. Faucett and D. C. Thomas. Simultaneously modelling censored survival data and repeatedly measured
covariates: A Gibbs sampling approach. \textit{Statistics in Medicine}, 15:1663-1685, 1996.
\item  A. K. Furgal, A. Sen, and M. Taylor, J. Review and comparison of computational approaches for joint
longitudinal and time-to-event models. International Statistical Review, 87(2):393-418, 2019.
\item  A. Gelman. Prior distributions for variance parameters in hierarchical models. \textit{Bayesian Analysis}, 1(3):
515-533, 2006.
\item  A. L. Gould, M. E. Boye, M. J. Crowther, J. G. Ibrahim, G. Quartey, S. Micallef f , and F. Y. Boisg. Joint
modeling of survival and longitudinal non-survival data: current methods and issues. Report of the DIA
Bayesian joint modeling working group. \textit{Statistics in Medicine}, 34:121-133, 2017.
\item  X. Guo and B. P. Carlin.Separate and Joint Modeling of Longitudinal and Event Time Data Using
Standard Computer Packages. \textit{The American Statistician}, 58(1):16-24, 2004.
\item  R. Henderson, P. Diggle, and A. Dobson. Joint modelling of longitudinal measurements and event time
data. \textit{Biostatistics}, 1(4):465-480, 2000.
\item  W. Hu, G. Li, and N. Li. A Bayesian approach to joint analysis of longitudinal measurements and competing
risks failure time data. \textit{Statistics in Medicine}, 28:1601-1619, 2009.
\item  Y. Huang and G. Dagne.A Bayesian Approach to Joint Mixed-Eff ects Models with a Skew-Normal
Distribution and Measurement Errors in Covariates. \textit{Biometrics}, 67:260-269, 2011.
\item  J. G. Ibrahim, M.-H. Chen, and D. Sinha. \textit{Bayesian Survival Analysis}. New York: Springer, 2001.
\item  J. G. Ibrahim, M.-H. Chen, D. Zhang, and D. Sinha. Bayesian Analysis of the Cox model. In J. P. Klein,
H. C. van Houwelingen, J. G. Ibrahim, and T. H. Scheike, editors,\textit{ Handbook of Survival Analysis}, chapter 2,
pages 27-49. Chapman \& Hall/CRC, Boca Raton, 2014.
\item  M. K$\ddot{o}$hler, N. Umlauf, A. Beyerlein, C. Winkler, A.-G. Ziegler, and S. Greven. Flexible bayesian additive
joint models with an application to type 1 diabetes research. \textit{Biometrical Journal}, 59:1144-1165, 2016.
\item  Z. Li, V. M. Chinchilli, and M. Wang. A Bayesian joint model of recurrent events and a terminal event.
\textit{Biometrical Journal}, 61:187-202, 2019.
\item  S. Luo. A Bayesian approach to joint analysis of multivariate longitudinal data and parametric accelerated
failure time. \textit{Statistical Methods in Medical Research}, 33:580-594, 2014.
\item  R. Martins, G. L. Silva, and V. Andreozzib. A Bayesian approach to joint analysis of multivariate longitu-
dinal data and parametric accelerated failure time. \textit{Statistics in Medicine}, 35(19):3368-3384, 2016.
\item  J. V. Niekerk, H. Bakka, and H. Rue. Joint models as latent gaussian models - not reinventing the wheel.
arxiv:1901.09365v1, 2019.
\item  C. Proust-Lima, M. S\`ene, J. M. G. Taylor, and H. Jacqmin-Gadda. Joint latent class models for longitudinal
and time-to-event data: A review. \textit{Statistical Methods in Medical Research}, 23:74-90, 2014.
\item  D. Rizopoulos. \textit{Joint models for longitudinal and time-to-event data}. New York: Chapman and Hall/CRC,
2012.
\item  D. Rizopoulos. The R package JMbayes for f i tting joint models for longitudinal and time-to-event data
using mcmc. \textit{Journal of Statistical Software}, 72(7):1-45, 2016.
\item  A.-M. Tang and N.-S. Tang. Semiparametric bayesian inference on skew-normal joint modeling of multi-
variate longitudinal and survival data. \textit{Statistics in Medicine}, 34(5):824-843, 2015.
\item  Y. Tian, E. Li, and M. Tian. Bayesian joint quantile regression for mixed ef f ects models with censoring
and errors in covariates. \textit{Computational Statistics}, 31:1031-1057, 2016.
\item  N. Umlauf, N. Klein, and A. Zeileis. BAMLSS: Bayesian additive models for location, scale and shape (and
beyond). \textit{Journal of Computational and Graphical Statistics}, 27(3):612-627, 2018.
\item  N. Umlauf, N. Klein, T. Simon, and A. Zeileis. bamlss: A Lego toolbox for flexible Bayesian regression
(and beyond), arxiv 1909.12345, 2019.
\item  G. Verbeke and M. Davidian. Joint models for longitudinal data: Introduction and overview. In G. Fitz-
maurice, M. Davidian, V. G., and M. G., editors, \textit{Longitudinal Data Analysis}, chapter 13, pages 319-326.
Chapman \& Hall/CRC, Boca Raton, 2009.
\item  Y. Wang and J. M. G. Taylor. Jointly Modeling Longitudinal and Event Time Data With Application to
Acquired Immunodeficiency Syndrome.\textit{ Journal of the American Statistical Association}, 96(455):895-905,
2001.
\item  W. Ye and M. Yu. Joint models of longitudinal and survival data. In J. P. Klein, H. C. van Houwelingen,
J. G. Ibrahim, and T. H. Scheike, editors, \textit{Handbook of Survival Analysis}, chapter 26, pages 523-548.
Chapman \& Hall/CRC, Boca Raton, 2014.
\item  M. Yu, N. Law, J. Taylor, and H. Sandler. Joint longitudinal survival-cure models and their application to
prostate cancer. \textit{Statistica Sinica}, 14(4):835-862, 2004.
\item  D. Zhang, M. H. Chen, J. G. Ibrahim, M. E. Boye, P. Wang, and W. Shen. Assessing model f i t in joint
models of longitudinal and survival data with applications to cancer clinical trials. \textit{Statistics in Medicine},
33:4715-4733, 2014.
\item  D. Zhang, M. H. Chen, J. G. Ibrahim, M. E. Boye, and W. Shen. Bayesian Model Assessment in Joint
Modeling of Longitudinal and Survival Data with Applications to Cancer Clinical Trials. \textit{Journal of
Computational and Graphical Statistics}, 26(1):121-133, 2017.
\item  H. Zhu, J. Ibrahim, Y. Y. Chi, and N. Tang. Bayesian inf l uence measures for joint models for longitudinal
and survival data. \textit{Biometrics}, 68(3):954-964, 2012.
\item  H. Zhu, S. M. DeSantis, and S. Luo. Joint modeling of longitudinal zero-inf l ated count and time-to-event
data: A Bayesian perspective. \textit{Statistical Methods in Medical Research}, 27(4):1258-1270, 2018.
\end{enumerate}

\end{document}